# THE RATE OF SUPERNOVAE: BIASES AND UNCERTAINTIES

ENRICO CAPPELLARO & MASSIMO TURATTO
*Osservatorio Astronomico di Padova*
*vicolo dell'Osservatorio 5, Padova, I-35122, Italy*

**Abstract.** Using an updated SN list and galaxy parameters from the RC3 catalog, we have revisited two well know biases affecting SN searches. We show that the bias in the central region of galaxies is negligible for galaxies closer than 40 Mpc (H=75 km s$^{-1}$ Mpc$^{-1}$), but causes the loss of almost 50% of SNe in the distance range 80-160 Mpc. This bias has a weak dependence on the galaxy and SN types. Instead the bias due to the inclination of spiral galaxies is more severe especially for SNII in late spirals with only 1 SNe out of 5 being detected in edge–on galaxies. The two effects seem interconnected with a more severe loss of SNe in the central part of inclined spirals.

Our best estimates of the SN rates are reported along with a quantitative estimates of the different source of errors.

## 1. Introduction

The current estimates of the rate of SNe still bring large uncertainties. This is due both to the small statistics, especially when considering a particular SN or galaxy type, and to the uncertainties in the corrections for the discovery bias affecting SN searches.

In particular, two such biases have been claimed to be especially important:

**a)** the discovery rate of SNe appears reduced in the central parts of more distant galaxies (Shaw, 1979)

**b)** in spiral galaxies the discovery rate has a strong dependence on the inclination of the galaxies (Tammann, 1974)

In the following we revisit the observational evidences of these two selection effects and investigate their dependence on galaxy and SN parameters.

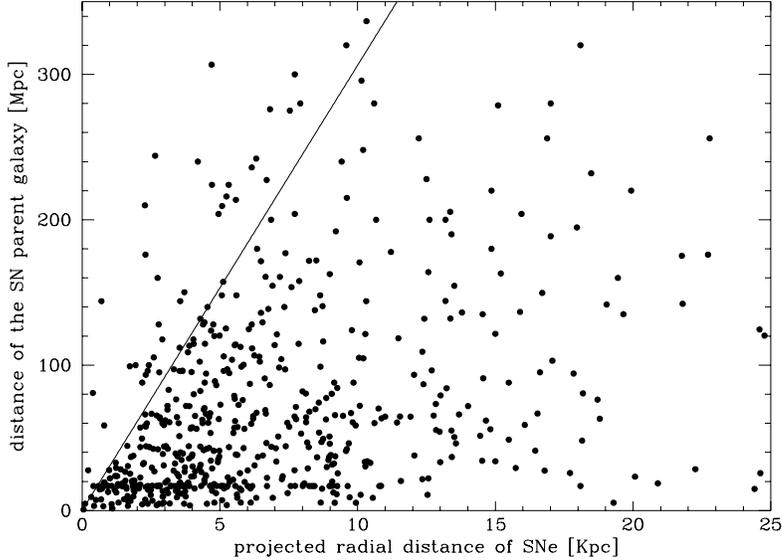

*Figure 1.* Distances of SN parent galaxies versus projected distances of SNe from the galaxy centers (H=75 km s$^{-1}$ Mpc$^{-1}$). 25 out of the total sample of 595 SNe fall beyond the limits of the graph. The line corresponds to an angular radial distance of 10$''$.

## 2. Discovery of SNe in the inner regions of galaxies

In Fig. 1 we plot the distance of SN parent galaxies versus the projected distances of the SNe from the galaxy centers.

The SN sample was extracted from the *Asiago Supernova Catalog* (Barbon *et al.*, 1989) updated to December 1994 and galaxy distances were derived from the radial velocities (de Vaucoluleurs *et al.* 1991 [RC3]), adopting H$_0$=75 km s$^{-1}$ Mpc$^{-1}$, except when these were smaller then 1500 km s$^{-1}$. In these cases the distances were retrieved from Tully (1988).

The figure is the same as Fig. 1 of Shaw (1979) but for the SN sample: excluding the SNe for which the galaxy recession velocities are not available, counts now 595 instead of 231 SNe. It is clear from Fig. 1 that in distant galaxies there is a deficiency of SNe at small radial distances. This is better seen in Fig.2 where we compare the distributions of SN radial distances in galaxies at different distances (this revises Fig.2 of Shaw 1979). In each panel the dotted line indicates the position of the median of the distribution: the fact that, increasing the distance, the median moves progressively to outer radii has to be interpreted as a selection effects in SN searches.

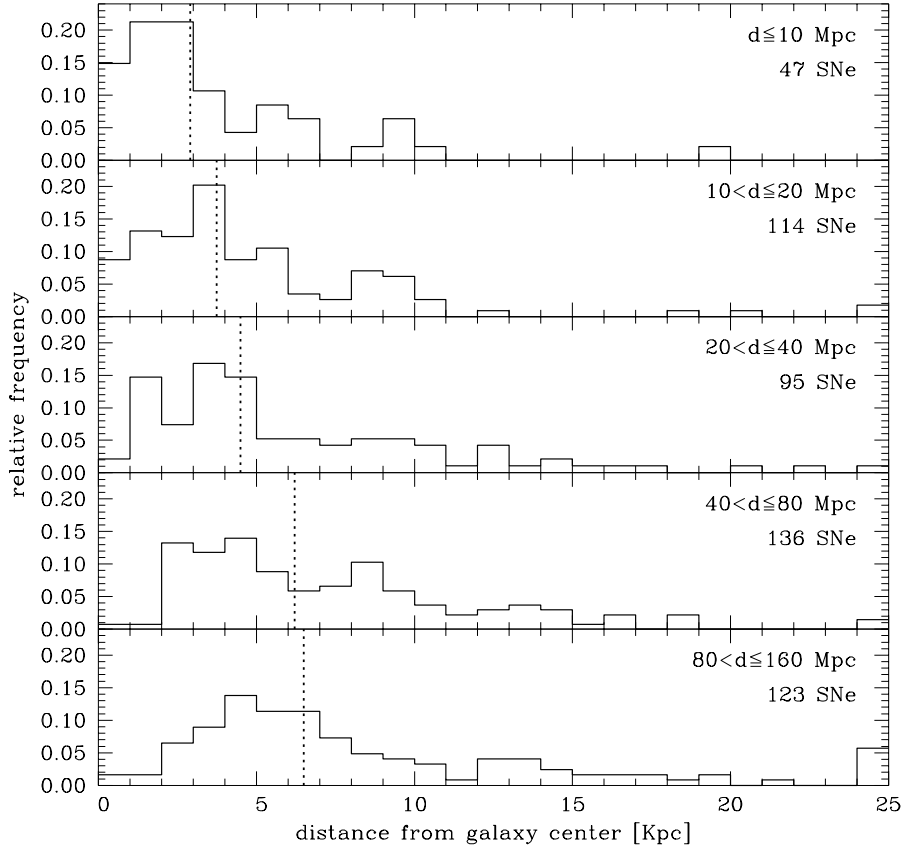

Figure 2. Distributions of SN radial distances in parent galaxies at different distances. In each panel the dotted line indicates the position of the median.

To estimate how many SNe remain undiscovered because of this effect, we assume that the SN sample in the nearest galaxies gives the intrinsic radial distribution and that, even for the more distant galaxies, the radial distributions are not biased outside a given radius $r_n$. The results depend both on the adopted distance bins and on the choice of $r_n$, the radii for the normalization. This is shown in Fig. 3 where is given the percentage of SNe lost as a function of the distance obtained using different recipes. In the figure, the dotted line (case $a$) gives the percentage of all SNe which are lost adopting the same bins and $r_n$ as in Shaw (continuous line). Despite the SN sample doubles that used by Shaw (1979), the results are similar. However, because of a better statistics, we can restrict the distance limit for the near reference galaxies to 10 Mpc instead of 22 Mpc. Also, to deal with the problem that few SNe have been discovered at large radial distances in nearby galaxies we choose to compare each distribution to the previous one (according to the distance) adopting for $r_n$ the radius corresponding to the median. The short-dash line in Fig. 2 gives the estimate of the percentage of SNe lost calculated according to this last recipe (case $b$).

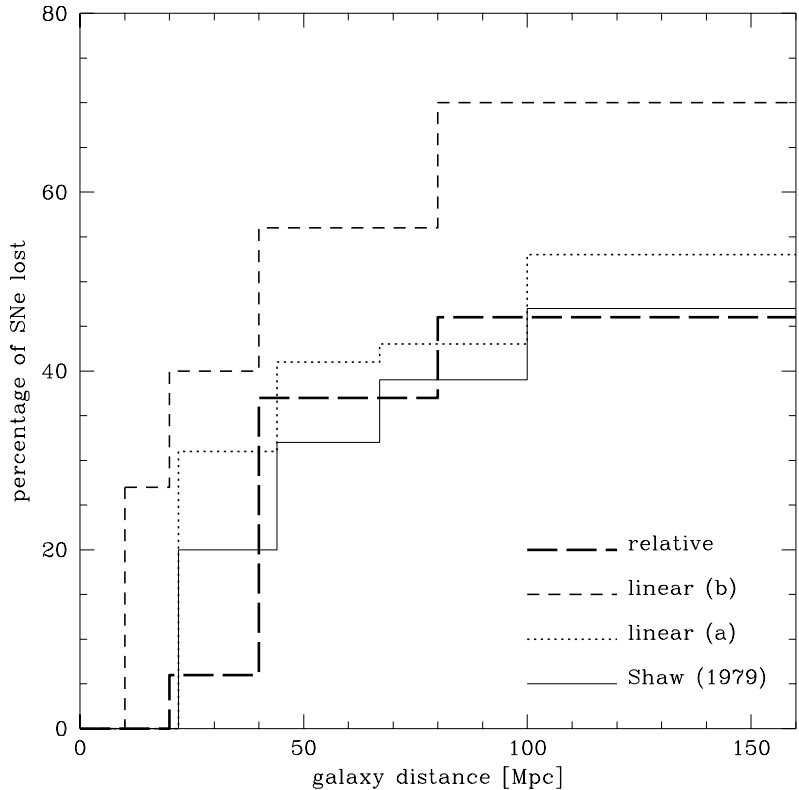

*Figure 3.*   Percentage of all SNe which are lost in the central parts of galaxies.

The fact that the bias appears strongly enhanced compared to case $a$ is due to the restriction in distance of the reference bin and seems to indicate that the computed estimates are lower limits for the true values (as also stressed by Shaw 1979). Taken to face value we should conclude that at least 70% of all the SNe which exploded in galaxies from 80 to 160 Mpc remained undiscovered.

A caveat that must be taken into account however is that the absolute radial distances of SNe in distant galaxies could be larger in average than in nearby galaxies only because distant parent galaxies are intrinsically brighter and bigger. This might be a consequence of the fact that the galaxy samples are intrinsically biased toward such galaxies.

TABLE 1. Percentage of SNe lost in the inner regions of galaxies in the range 40-160 Mpc with respect to those discovered in galaxies closer than 40 Mpc. Relative radial distributions have been compared.

|          | All     | E-S0    | S0a-Sb  | Sbc-Sd  | $i \leq 45°$ | $i > 45°$ | Ia      | II+Ib/c |
|----------|---------|---------|---------|---------|--------------|-----------|---------|---------|
| % lost SNe | $35 \pm 8$ | $23 \pm 12$ | $36 \pm 13$ | $30 \pm 8$ | $19 \pm 6$   | $43 \pm 12$ | $27 \pm 10$ | $38 \pm 15$ |
| no. SNe  | 444     | 55      | 119     | 214     | 155          | 203       | 96      | 143     |

To test this possibility we computed the percentage of SN lost using, relative radial distances, that is the linear radial distances have been normalized to the galaxy radii (using $d_{25}$ as reported in RC3). Again the SN parent galaxies are divided in five distance bins (as in Fig.2) and the percentage of SNe lost is computed assuming that the distributions at radii larger than those corresponding to the median are unbiased: the results are shown in Fig.3 as a long-dashed line. By using the relative radial distances the central bias results smaller than that obtained with linear radial distance appearing negligible up to 40 Mpc while in the range 80-160 Mpc about 50% of all SNe are lost. This proves that distant SN parent galaxies are in the average bigger than nearby ones.

In Table 1 we report the estimates of the percentages of SNe lost for different classes of galaxies and SNe. In order to get a better statistics only two broad distance bins have been considered: the relative radial distribution of SNe in galaxies in the range 40-160 Mpc has been compared to that of galaxies closer than 40 Mpc.

Based on this table we conclude that:
1. there are not significant differences between early and late type galaxies but statistics for early type galaxies is poor.
2. dividing the spiral galaxy sample in two classes according to the inclination ($i$) of the disk with respect to the line of sight, we find that the fraction of SNe lost in the central region of inclined ($i > 45°$) spirals is twice as large as that in less inclined ($i \leq 45°$) galaxies. Therefore, there is a interplay of the central bias with the inclination effect.
3. the effect appears stronger for SNe with massive progenitor, i.e. type II+Ib/c, than for SN Ia as also suggested, but not quantified, by van den Bergh (1994).

Concerning the cause of this observational bias we believe that it is related to the area of confusion around the bright galaxy nuclei which de-pends on the scale of the telescope, detector type, etc. For the wide field telescopes usually employed in SN searches this area has an angular dimension of several arcseconds. For instance, for the Asiago 67/92cm Schmidt we verified that the confusion radius is of the order of 10 arcsec (Turatto et al., 1994). It is suggestive to note (Fig. 1) that the line corresponding to an angular radius of 10 arcsec gives a reasonable inner boundary for the discovery of SNe in galaxies at different distances.

## 3. Discovery of SNe in inclined galaxies

The presence of a bias on the discovery of SNe in inclined, late–type spirals was first pointed out by Tammann (1974). Based on the SNe discovered at that time in a sample of nearby galaxies (50 SNe) he found that "the nearly face-on Sc's give a rate which is $6.4 \pm 1.8$ times higher" than the rate in inclined ones. The effect was confirmed by van den Bergh and McClure (1990) again on a nearby galaxy sample (59 SNe) and by Cappellaro et al. (1993b) on the data of the Asiago+Crimea SN search (56 SNe). In the last paper it was also suggested that the effect is stronger in late spirals and for SNe with massive progenitors (types II and Ib). On the other side there are indications that CCD and visual searches are not affected (Muller et al., 1992; van den Bergh & McClure, 1994).

Whether the correction for this bias is included or not, makes a large difference especially for the estimate of the rate of SNII in late spirals (a factor 3 on the overall rate according to Tab.6 of Cappellaro et al. 1993b).

To verify the importance of the inclination effect on the list of SNe discovered to December 1994 we divided the spiral parent galaxies in three classes according to the inclination of the disk along the line of sight. Inclination was calculated based on the axial ratio, $R_{25}$, as given by RC3 catalog, using the formula:

$$i = \arccos \sqrt{(10^{-2R_{25}} - 0.04)/0.96}$$

To reduce the interference with the central bias discussed in the previous section, we selected only galaxies with $d \leq 40$ Mpc (the sample counts 222 SNe). The number of SNe in galaxies at different inclination have been normalized by comparison with the distribution of inclination of spirals in the RC3 catalog and the percentage of SNe which are lost have been calculated assuming that all SNe in face on galaxies have been discovered.

We confirm that the inclination bias is severe. In the whole, 50% of all SNe are lost because of this effect. The problem is worsened by the fact that, because the bias is different for different kinds of galaxies and SNe, also the relative rates are heavily affected.

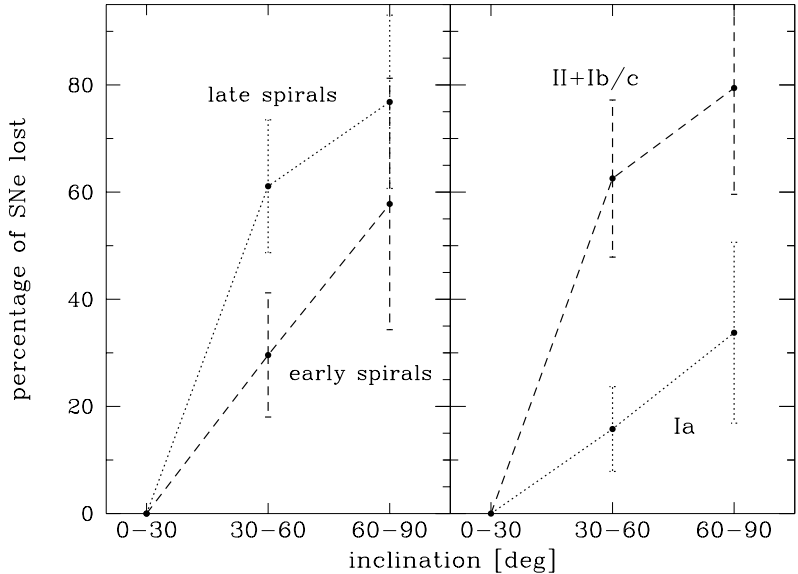

*Figure 4.* Percentage of SNe which are lost in inclined spirals with respect to face on galaxies. In the left panel, parent galaxies are divided in early and late type spirals, in the right panel they are divided according to the SN type. Only galaxies with $d \leq 40$ Mpc are included. Statistical error bars are indicated.

We illustrate this point in Fig.4, where in the left panel parent galaxies are divided in early (S0a-Sb) and late (Sbc-Sd) spirals while in the right panel they are divided according to the SN type (galaxies with $n$ SNe are counted $n$ times). The bias results most severe for SNe II+Ib/c and in late spirals in which 4 out of 5 SNe are missed when the inclination is large, less severe for SN Ia and in early spirals, where only 1 out of three SNe is lost.

Concerning the origin of the inclination bias, more likely there are two concurrent causes. One cause is certainly the extinction by dust in the disk of spirals. Since most SNe belong either to the young stellar arm population (II, Ib/c) or to an intermediate age disk population (Ia) the average extinction is expected to increase with the inclination of the galaxy and, as a consequence, the chance to discover a SN decreases. This interpretation explains why the bias is stronger in late spirals, which contain more dust, than in early ones (Fig. 4, left panel). However, at least for a uniform dust distribution (cf. van den Bergh 1990), the bias is expected to be severe only in almost edge on spirals. Instead we found a strong effect also in moderately inclined galaxies.

A second cause of the inclination bias is likely a photographic effect. The average density of the images of inclined galaxies on photographic plates is higher than that of face-on galaxies, increasing the "area of confusion". This hampers the discovery of SNe in inclined galaxies in a similar way as for the bias in the central region of galaxies and can explains the interconnection between the two biases.

Visual and CCD searches are expected to be less affected by both the afore mentioned problems. Indeed we reminded earlier that they have been claimed not to suffer for the inclination bias.

The fact that the inclination effect is much more severe for SNe II+Ib/c than for SN Ia (Fig. 4 right panel) fits with this interpretation. SNe II and Ib/c are confined to regions of recent star formation, close to spiral arms and young star associations which give a higher background density. Also these are dust rich regions in which extinction is expected to be important. In addition, SNe II and Ib/c are intrinsically fainter than SN Ia and the contrast between a SN II or Ib/c event and the local background is significantly less than in SN Ia.

## 4. Estimate of the errors in the SN rates

The SN rate estimates reported in the literature do not always includes the error estimates. When they do, usually they refer only to the statistical errors calculated assuming a Poisson statistics for the SN events. It is true that statistical errors dominate when the SN sample is small, but there are at least two additional sources of errors:

1. we have shown in the previous sections that biases on the SN discovery may be severe. Therefore, it is necessary to account properly for them. Corrections depend on several factors, namely the SN search characteristic, the distances, morphological types and inclinations of the parent galaxies, the SN types. For this reason there are large uncertainties on the correction factors to be applied.
2. depending on the approach, there are a number of assumptions that must be made for the calculation of the SN rate. For instance, if the "fiducial sample" method is used (Tammann *et al.*, 1994) one needs to assume that all galaxy in the fiducial sample have been searched for SNe at an equal intensity level independently on galaxy and SN type. Unfortunately, there are evidences that this assumption does not hold for present SN samples (Turatto *et al.*, 1994), but it is difficult to quantify the errors. Instead, in the calculations of the SN rate with the "control time" method a number of input parameters are involved each with an estimated error. In particular, Cappellaro et al.(1993a,1993b) investigated the propagation of the errors on the SN average absolute magnitudes and dispersions, the SN light curve shapes and the limiting magnitude of the telescope(s) used for the search(es). In this way it has been possible to give a quantitative estimate of the errors.

TABLE 2. Relative errors on SN rates derived from Tab.7 of Cappellaro et al. 1993b. The total errors have been obtained by adding quadratically the individual components.

| galaxy type | SN statistics | | | input parameters | | | correction factors | | | total errors | | |
|---|---|---|---|---|---|---|---|---|---|---|---|---|
| | Ia | Ib/c | II | Ia | Ib/c | II | Ia | Ib/c | II | Ia | Ib/c | II |
| E-S0 | 0.38 | | | 0.15 | | | 0.15 | | | 0.44 | | |
| S0a-Sb | 0.29 | 0.69 | 0.47 | 0.12 | 0.23 | 0.17 | 0.29 | 0.38 | 0.40 | 0.43 | 0.82 | 0.63 |
| Sbc-Sd | 0.26 | 0.55 | 0.24 | 0.13 | 0.19 | 0.18 | 0.38 | 0.33 | 0.32 | 0.48 | 0.67 | 0.44 |

In Tab.2 we report the estimates of the relative errors on SN rates for the different SN and galaxy types. Whereas they have been calculated for the combined Asiago+Crimea SN searches (Cappellaro et al., 1993b) they are expected to give the order of magnitude of the errors also for other recently published estimates of the SN rate.

It can be noted that a) errors in the input parameters are never dominant, b) SN statistics errors dominates in early type galaxies and for SN Ib/c and c) errors on the corrections factors are especially important for SN II in late spirals.

## 5. Conclusions

With the aim to obtain more reliable determinations of the frequency of SNe we are moving in two directions. On one side we try to improve the SN statistics by merging the databases of different SN searches, still with proper account for the individual characteristics of each search. On the other side we analyze in detail the selection biases on SN searches.

In this paper we have discussed the bias on the SN discoveries in the central regions of the distant parent galaxies and the selection effect due to inclination of spirals. The first bias has been revisited with a sample of SNe twice as large as in Shaw (1979). We find that for galaxies in the distance range 80-160 Mpc about 50% of all SNe are lost due to the inability of the searches to discover SNe in the central regions of the galaxies. The effect is found to depends only weakly on the galaxy and SN type. The inclination effect, on the contrary, has strong dependence on the galaxy morphology and SN type. We find that in inclined ($60 < i \leq 90$) late spirals only 1 out of 5 SNII are discovered. We noted also that the two effects are not

10     ENRICO CAPPELLARO & MASSIMO TURATTO

TABLE 3. SN rates in SNu (SNe $\times\, 10^{10}\, L_\odot^{-1} \times (100\text{yr})^{-1}$. Rates in SNu scales as $(H/75)^2$.

| galaxy | Ia | Ib/c | II |
|---|---|---|---|
| E-S0 | $0.13 \pm 0.06$ | $\leq 0.06$ | |
| S0a-Sb | $0.17 \pm 0.07$ | $0.13 \pm 0.11$ | $0.30 \pm 0.19$ |
| Sbc-Sd | $0.39 \pm 0.19$ | $0.27 \pm 0.18$ | $1.48 \pm 0.65$ |

independent in the sense that the deficiency of SNe in the central regions of distant galaxies is more pronounced for more inclined galaxies.

In conclusion we report in Tab.3 our current best estimates of the SN rates along with the errors. These have been derived applying the control time method to the combined logs of the Asiago and Crimea searches (Cappellaro et al. 1993a,1993b).

The data relative to external galaxies can be used also to obtain an estimate of the rate of SNe in the Galaxy by assuming that its morphological type is Sb±0.5 and the luminosity is $L_B = 2.0 \pm 0.6 \times 10^{10}\, L_\odot$. With these assumption we predict in the Galaxy $3 \pm 2$ SN Ia, $2 \pm 2$ SN Ib and $12 \pm 8$ SN II per millennium. Given the large errors, which however are of the same order of those of other methods, the predicted galactic SN rates are in good agreement with other observational evidences, e.g. those of historical SNe and SN remnants, and with theoretical expectations.